\documentclass[twocolumn,superscriptaddress,altaffilletter, showkeys, 10pt, aps,prd]{revtex4-2}
\usepackage{amssymb}
\usepackage{amsfonts}
 
\usepackage{graphicx}

\usepackage{enumerate}

\usepackage{xcolor}
\usepackage{amsmath}
\usepackage{amstext}
 \usepackage[dvipdf]{epsfig}
  \usepackage{color}
\def\be{\begin{equation}}
\def\ee#1{\label{#1}\end{equation}}
\newcommand{\ben}{\begin{eqnarray}}
\newcommand{\een}{\end{eqnarray}}

\begin{document}
\title{Quasinormal modes and the analytical continuation of non-self-adjoint operators}

\author{Mart\'{\i}n G. Richarte}\email{martin@df.uba.ar  }
\affiliation{PPGCosmo, CCE - Universidade Federal do Esp\'irito Santo,  29075-910 Vit\'oria, ES, Brazil}
\affiliation{Departamento de F\'isica, Facultad de Ciencias Exactas y Naturales,
Universidad de Buenos Aires, Ciudad Universitaria 1428, Pabell\'on I,  Buenos Aires, Argentina}
\author{J\'ulio C. Fabris}\email{julio.fabris@cosmo-ufes.org}
\affiliation{PPGCosmo, CCE - Universidade Federal do Esp\'irito Santo,  29075-910 Vit\'oria, ES, Brazil}
\affiliation{N\'ucleo Cosmo-ufes\&Departamento de F\'isica - Universidade Federal do Esp\'irito Santo,   29075-910  Vit\'oria, ES, Brazil}
\author{Alberto Saa}\email{asaa@ime.unicamp.br}
\affiliation{Departamento de Matem\'atica Aplicada, Universidade Estadual de Campinas, 13083-859 Campinas, SP, Brazil}
\bibliographystyle{plain}

\begin{abstract}
We briefly review the analytical continuation method for determining quasinormal modes (QNMs) and the associated frequencies in open systems. We explore two exactly solvable cases based on the P\"oschl-Teller potential to show that the analytical continuation method cannot determine the full set of QNMs and frequencies of a given problem starting from the associated bound state problem in Quantum Mechanics. The root of the problem is that many QNMs are the analytically continued counterparts of solutions that do not belong to the domain where the associated Schr\"odinger operator is self-adjoint, challenging the application of the method for determining full sets of QNMs. We illustrate these problems through the physically relevant case of BTZ black holes, where the natural domain of the problem is the negative real line.  
\end{abstract}

\vskip 2cm

\date{\today}
\maketitle

\section{Introduction}

The QNM analysis is one of the main strategies used to inspect the stability of many physical open systems,  with many applications ranging from optics to General Relativity\cite{Review0,Review,Review1}. In their simplest formulation, QNM are separable solutions  
\begin{equation}
\label{QNM}
 \Psi(t,x) = e^{-i\omega t} u(x) 
\end{equation}
of an $(1+1)$-dimensional wave equation.  After a separation of variables procedure,  $u(x)$ is typically expected to obey  a  Schr\"odinger-like second-order linear differential  equation 
\begin{equation}
\label{qnm}
\left(- \frac{d^2}{dx^2} + V(x)\right) u = \omega^2 u 
\end{equation}
on a certain domain of  $\mathbb{R}$.
For situations where the modes $u$ are defined on the entire real line $\mathbb{R}$ and the potential $V(x)$  vanishes sufficiently fast for $x\to\pm\infty$, the QNM frequencies are defined as the (typically complex) values of $\omega$ such that the solutions of (\ref{qnm}) behave as outgoing waves at   $x\to \infty$  and ingoing
ones at  $x\to -\infty$, corresponding intuitively to solutions that disperse towards  infinity. According to our definition for $\Psi$, these outgoing/ingoing waves correspond,  respectively, to solutions of (\ref{qnm})
such that 
\begin{equation}
\label{plus-inf}
u \propto e^{  {i\omega x}}, \quad {\rm for}  \quad x\to\infty,
\end{equation}
and
\begin{equation}
\label{minus-inf} 
  u \propto e^{- {i\omega x} } ,  \quad {\rm for}  \quad \to-\infty.
 \end{equation}
Since (\ref{qnm}) admits as solutions both $\omega$ and $-\omega$, we need to
 assume here $\Re~(\omega) \ge 0$, otherwise the QNMs are not unambiguously defined. 
Also according to our definition, the modes will be exponentially suppressed in time if $\Im ~(\omega) < 0$. Notice that, in contrast with
the usual spectral theory of Schr\"odinger operators in Quantum Mechanics, the eigenvalues $\omega^2$ in (\ref{qnm}) can be, and usually are, complex and the QNMs are not, in general, a complete set for the problem \cite{Review0}. 
 
 In standard situations involving asymptotically flat black holes  in General Relativity (see, for references, \cite{Review, Review1}),   the equivalent of equation (\ref{qnm}) is obtained by introducing some sort of radial tortoise-coordinate $x$ in the exterior region of the black hole, and typically in these cases the effective potential $V(x)$ is non negative and  has a barrier shape. Moreover,   the conditions (\ref{plus-inf}) and  (\ref{minus-inf}) have the usual interpretation of wave solutions escaping to infinity and plunging into the event horizon, respectively, implying that  QNMs are always associated with dispersive phenomena for these systems since they imply a net transport of energy towards outside the system. 

In the present paper, we will review the analytical continuation method for determining QNMs and frequencies for problems of the type (\ref{qnm}), starting from an associated bound state problem in Quantum Mechanics. Through two explicit examples based on exactly solvable P\"oschl-Teller potentials, we will show that the analytical continuation method cannot determine the complete set of QNMs and that the origin of the problem is that QNMs are typically the analytically continued counterparts of solutions that belong to domains where the associated Schr\"odinger operator fails to be self-adjoint.

\section{Analytical continuation of  Schr\"odinger operators}

It is rather common to compute the QNMs and their associate frequencies $\omega $ for the equation (\ref{qnm}) with a given potential barrier $V$ through a formal analytical continuation performed in the bound state  problem of a Schr\"odinger operator $\mathcal{H}$ associated with the potential well corresponding to the inverted
 potential $\tilde V = -V$.  Such an approach, introduced decades ago by Blome, Ferrari, and Mashhoon \cite{ba1,ba2,ba3},  is one of the best options we have at hand to obtain analytical answers and gain some physical insights on the QNM problem. The approach consists basically in a formal map between the QNM solutions of (\ref{qnm}) and the bound  states of the quantum mechanical problem governed by the Schr\"odinger  operator 
\begin{equation}
\label{schroed}
{\cal H}\psi = \left( -\frac{\hbar^2}{2m}\frac{d^2}{dx^2} + \tilde V(x)\right)\psi = E\psi .
\end{equation}
We know that  for $\tilde V(x)$  vanishing  sufficiently fast for $x\to\pm\infty$, 
the bound states of $\cal H$ will decay exponentially, {\em i.e.},
\begin{equation}
\label{Hplus-inf}
\psi \propto e^{- \sqrt{-\frac{2mE}{\hbar^2}}x  }, \quad {\rm for}  \quad x\to\infty,
\end{equation}
and
\begin{equation}
\label{Hminus-inf} 
\psi \propto e^{  \sqrt{-\frac{2mE}{\hbar^2}}x   } ,  \quad {\rm for}  \quad \to-\infty.
 \end{equation}
 Since the literature on bound states of Schr\"odinger operators is huge, with many studies exploring a vast range of different potentials, this method is commonly beneficial for identifying exact or approximate QNMs.
  
 The original approach is based on the extension of the solutions of (\ref{qnm}) or (\ref{schroed}) for the entire complex plane by means
 of the formal substitution (Wick rotation) $x\to ix$, which reduces the QNM boundary conditions 
(\ref{plus-inf}) and (\ref{minus-inf}) 
 to the bound  state ones (\ref{Hplus-inf}) and (\ref{Hminus-inf}). After some parameter redefinitions in the potential $V(x)$, one can effectively map the QNMs on the bound  states of (\ref{schroed}), and, consequently, relate the QNM frequencies $\omega$
 of (\ref{qnm}) with the energy spectrum $E$ of
 $\mathcal{H}$. More explicitly, suppose we know a bound state $\psi$ of (\ref{schroed}). It should have an associate
 eigenvalue (energy) $E<0$, since $\tilde V$ is assumed to be a non-positive potential well. Suppose
 also that the potential $\tilde V$ depends on a set of real parameters $\alpha_k$, $k=1,2,\dots$, $\tilde V = \tilde V(x,\alpha_k)$. Clearly, both the eigenfunction $\psi$ and the energy $E$ may have similar dependence on the parameters, {\em i.e.},
$\psi=\psi(x,\alpha_k)$ and   $E=E(\alpha_k)$. 
 After the formal substitution 
 $x\to -ix$, the Schr\"odinger equation (\ref{schroed}) will read
 \begin{equation}
 \label{schroed1}
 \left( - \frac{d^2}{dx^2} - \frac{2m}{h^2 }\tilde V(-ix,\alpha_k)\right)\psi  = - E(\alpha_k)\psi ,
 \end{equation}
 and the asymptotic conditions (\ref{Hplus-inf}) and (\ref{Hminus-inf}) for $\psi$ are formally transformed in
 (\ref{plus-inf}) and (\ref{minus-inf}) for $\psi(-ix)$.
Suppose  now  we can  transform the parameter $\alpha_k$ in such way that the potential $\tilde V$ remains invariant under the Wick rotation, {\em i.e.}, let us introduce a new set of parameters $\alpha_k'$ such that
 \begin{equation}
 \tilde V(x,\alpha_k) = \tilde V(-ix,\alpha_k').
 \end{equation}
 With this transformation, equation (\ref{schroed1}) will read
 \begin{equation}
 \label{qnm1}
  \left( - \frac{d^2}{dx^2} -  \tilde V(x,\alpha_k)\right)u = - E(\alpha_k')u ,
 \end{equation}
 with $u(x)=\psi(-ix,\alpha_k')$. For sake of simplicity, we have set $\frac{h^2 }{2m} = 1$, without generality loss. Comparing (\ref{qnm1}) with (\ref{qnm}), we see that $u(x)$ is a QNM of the barrier potential corresponding to the inverted potential well $\tilde V$ with QNM frequency $\omega$ such that
 \begin{equation}
 \omega^2 = - E(\alpha_k'). 
 \end{equation}
 
  This method was sensibly simplified by the   prescription introduced
recently by Hatsuda \cite{borel}, which  is based in the following observation. Let us consider the
Schr\"odinger operator
\begin{equation}
\label{schroed2}
{\cal H}_\epsilon\psi = \left( -\epsilon^2\frac{d^2}{dx^2} + \tilde V(x)\right)\psi = E_\epsilon\psi ,
\end{equation}
where $\tilde V$ is a well behaved potential well in the entire real line $\mathbb{R}$ and $\epsilon>0$ is some
typical scale of the problem. Suppose $\psi_\epsilon(x)$ is a bound state
of ${\cal H}_\epsilon$ with energy $E_\epsilon$.
Consider now the analytical continuation of the Schr\"odinger operator given by ${\cal H}_{i\epsilon}$. The function  $u_\epsilon = \psi_{i\epsilon}$ is a QNM of the inverted potential $-\tilde V$, with frequency given by $\omega^2_\epsilon = -E_{i\epsilon}$.

 Before we consider the physically relevant case of BTZ black holes, let us consider a simple explicit example to illustrate better the analytical continuation method.

\subsection{The P\"oschl-Teller potential well}

 The so-called 
 P\"oschl-Teller potentials\cite{PT}
were the very  first family of non-elementary exactly soluble potential in Quantum Mechanics. We will illustrate 
the analytical continuation method with the P\"oschl-Teller potential corresponding to 
potential well  defined for the entire real line $\mathbb{R}$
\begin{equation}
\label{PTP}
\tilde V (x)= -  \frac{V_0}{\cosh^2 x},
\end{equation}
The Schr\"odinger equation (\ref{schroed1}) with this potential 
 admits bound states with energy spectrum  given by (see, for instance, \cite{Flugge})
\begin{equation}
E_\epsilon^{(n)}  = -\left(\sqrt{   { V_0}+\frac{\epsilon^2}{4}}-  \epsilon\left(n + \frac{1}{2}\right)  \right)^2,
\end{equation}
 with $n$ integer such that $0\le n \le n_{\rm max}$, where
\begin{equation}
n_{\rm max}  = \left\lfloor \frac{1}{2}\left(1+\sqrt{   \frac{4V_0}{\epsilon^2} + 1}\right) \right\rfloor.
\end{equation} 
It is important to stress that we have only a finite number of bound states for the P\"oschl-Teller potential  well. This is a well
known property in Quantum Mechanics for potential wells vanishing sufficiently fast for $x\to\pm\infty$.

We can now apply the Hatsuda prescription, and we will have the following set of QNM frequencies
\begin{equation}
\label{qnmfh}
\omega_\epsilon^{(n)}  =    \sqrt{   { V_0}-\frac{\epsilon^2}{4}}-  i\epsilon\left(n + \frac{1}{2}\right) 
\end{equation}
for the P\"oschl-Teller potential barrier $V=-\tilde V$.
However, one could solve exactly the QNM problem for the inverted P\"oschl-Teller potential  well $V$ (see, for instance, \cite{Review}),
and we would get the QNM frequencies (\ref{qnmfh}) without the restriction $0\le n \le n_{\rm max}$. In other words,
the  P\"oschl-Teller potential barrier has infinitely many QNM frequencies, and only a small set of them can be obtained
from the analytical continuation of the Schr\"odinger operator. If one reverses the analytical continuation procedure, we will have
that the QNM with $n>n_{\rm max}$ are mapped in solutions of the Schr\"odinger equation that do not correspond to bound states and, hence, do not belong the the usual domain where $\mathcal{H}_\epsilon$ is self-adjoint. This simple example shows that one cannot get the
full set of QNM frequencies starting from the bound states of the associate Quantum Mechanics problem. 
Notwithstanding, the P\"oschl-Teller potential  is effectively used to compute some QNMs in the space-times of black holes as far as they can mimetize the effective potential in the vicinity of the horizon. The results using P\"oschl-Teller potential can be compared with a numerical analysis, and the agreement is generally very good, the difference between both computations being less than $1\%$ and decreases as the effective potential becomes more localized, see Ref. \cite{molina}.

\section{ BTZ black holes}

 The BTZ black hole  \cite{btz} is an appealing  solution in three-dimensional gravity with a negative cosmological constant, $\Lambda=-1/\ell^{2}$. In  the case of zero angular momentum ($J=0$), its event horizon is determined solely by its mass $M$ and the AdS length scale, $\ell$. To begin with, we note that the line element for the exterior BTZ black hole with $J=0$ can be expressed as follows:
 \begin{equation}
ds^{2}=-\frac{r^{2}-r^{2}_{+}}{\ell^2}dt^{2}+ \frac{\ell^2}{r^{2}-r^{2}_{+}}dr^{2}+r^{2}d\theta^{2},
\end{equation}
where $t \in \mathbb{R}$, $r>r_{+}$, and $\theta \in [0,2\pi)$. In this context, the horizon can be expressed in terms of $\ell$ and $M$ as follows: $r^{2}_{+}=M \ell^{2}$ \cite{btz}, as previously noted.

We consider a massless Klein-Gordon  scalar field on this background, 
\begin{equation}\label{kg}
\Box\Phi=0.
\end{equation}
We express the scalar field by means of the parametrization $\Phi = e^{-i\omega t} e^{i \mu \theta} u(r)/\sqrt{r}$, where $\mu \in \mathbb{Z}$ and $\omega \in \mathbb{C}$, the latter representing the quasi-normal mode frequencies according with our definitions.  The case of a massive scalar field propagating on the rotating BTZ background can be found in \cite{bir}.

Considering the definition of the tortoise coordinate, expressed through the familiar relation $dx=dr/f(r)$, we arrive at the following expression:
\begin{equation}\label{tc}
x=-\frac{\ell^2}{r_{+}} \coth^{-1}\Big(\frac{r}{r_{+}}\Big).
\end{equation}
Eq. (\ref{tc}) tells us that the tortoise coordinate  effectively maps the interval  $(r_{+}, +\infty)$ onto $(-\infty, 0)$. 
By combining this result (\ref{tc}) with the equation outlined in (\ref{kg}), it led to a Schr\"odinger-like second-order linear differential equation,

\begin{equation}\label{qnmf}
\Big(-\frac{d^2}{dx^2} + V[r(x)]\Big)u=\omega^{2} u,
\end{equation}
where $f=\frac{r^{2}-r^{2}_{+}}{\ell^2}$, and the effective potential reads 

\begin{equation}\label{pot}
V= \frac{V_{0}}{\sinh^{2}(\alpha x)}+ \frac{V_{1}}{\cosh^{2}(\alpha x)}.
\end{equation}
Here, we define $\alpha=r_{+}/\ell^2$, $V_{0}=3\frac{r^{2}_{+}}{4\ell^{2}}>0$, and $V_{1}=\frac{r^{2}_{+}}{4\ell^{2}}(1+\frac{4\mu^2}{r^{2}_{+}})>0$. It is important to note that when $\mu = 0$, we return to the scenario examined in \cite{sunneta}. From this point onward, our goal will be to identify the QNMs associated with the equations given in (\ref{qnmf}) and (\ref{pot}). In this context, we will  analyze the boundary conditions pertinent to the half-real (negative) line. As is widely known, this generalized P\"oschl-Teller potential represents an exactly integrable problem, as established in \cite{Bin} and \cite{molina}.
Yet, the physical contexts differs significantly. The investigation of the QNMs for the pure de Sitter spacetime is addressed in \cite{Bin}, whereas the scattering problem associated with the generalized P\"oschl-Teller potential is thoroughly explored in \cite{molina}. The boundary conditions typically imposed at the horizon needs to be a purely incoming wave, represented as $e^{i\omega x}$, provided that a BTZ black hole is present. Conversely, at spatial infinity, we require an outgoing wave, $e^{-i\omega x}$, in order to eliminate any incoming radiation. However, the BTZ potential given in (\ref{pot}) approaches zero at the horizon while diverging as one moves toward infinity. For a solution to be well-defined near infinity, it must decay to zero. The specific cases wherein this decay condition is satisfied are what determine the QNMs frequencies \cite{molina}, \cite{nos}. 

After applying a new variable $z=\cosh^{-2}(\alpha x) \in [0,1)$  which compatifies the interval $\mathbb{R}_{-}$, the original master equation (\ref{qnmf}) can be recast as  the Gaussian hypergeometric equation \cite{nos}:
\begin{equation}
\label{gaussian}
z(1-z)u''~+ [c-(a+b+1)z]u'~ -ab u=0,
\end{equation}
where the parameter of the Gaussian hypergeometric are given by
\begin{eqnarray}
    a&=&\frac{1}{2}-i\frac{\omega}{2\alpha}+\frac{1}{4}\left(\nu+\zeta~\right),\\
     b&=&\frac{1}{2}-i\frac{\omega}{2\alpha}+\frac{1}{4}\left(\nu-\zeta~\right),\\
    c&=&1-i\frac{\omega}{\alpha}.
\end{eqnarray}

Here $\nu=\sqrt{1+4\frac{V_0}{\alpha^2}}$ and $\zeta=\sqrt{1-4\frac{V_1}{\alpha^2}}$. 
\begin{widetext}
Depending on the value of $c$, we can derive various types of solutions. Specifically, when $c \not\in \mathbb{Z}$, we find that the basis of linearly independent solutions is:   

\begin{eqnarray}
    &u&_{I}=z^{-i\frac{\omega}{2\alpha}}(1-z)^{\frac{1}{4}(1+\nu)} {}_2F_{1}(a,b,c, z),\\~~~~
    &u&_{II}=z^{+i\frac{\omega}{2\alpha}}(1-z)^{\frac{1}{4}(1+\nu)} {}_2F_{1}(a-c+1,b-c+1,2-c, z). ~~~~~~
\end{eqnarray}
\end{widetext}
At this stage, several comments are in order. When we consider the limit as $x \rightarrow -\infty$ and the fact that the hypergeometric function is equal to one when evaluated at the origin,  the boundary condition of having an ingoing-wave at the horizon implies that the second solution $u_{II}$ must be discarded. The other boundary condition corresponds to  impose that at infinity ($z\rightarrow 1^{-}$) the solution decays to zero, $\lim_{x\rightarrow 0}u_{I}=0$. To do so, we employ the Gursat's transformation to write ${}_2F_{1}(a, b, c, z)$ in terms of  a combination of ${}_2F_{1}(a, b, c, 1-z)$ \cite{Abramowitz}. Expanding $z=1-(\alpha x)^{2}+\mathcal{O}[(\alpha x)^{2}]$, the local expansion of the solution reads, 
\begin{equation}
\label{lim}
u_{I} \simeq  A(\alpha x)^{\frac{1}{4}(1+\nu) }  + B(\alpha x)^{\frac{1}{4}(1-\nu) },
\end{equation} 
with 
\begin{equation}
\label{A}
A = \frac{\Gamma(1-i\frac{\omega}{\alpha})\Gamma\left(-\frac{\nu}{\alpha}\right)}{\Gamma\left(\frac{1}{2}-i\frac{\omega}{2\alpha}-\frac{1}{4}\left(\nu+\zeta\right)
 \right) \Gamma\left(\frac{1}{2}-i\frac{\omega}{2\alpha}-\frac{1}{4}\left(\nu-\zeta\right)
 \right)}
\end{equation}
and
\begin{equation}
\label{B}
B = \frac{\Gamma(1-i\frac{\omega}{\alpha}) \Gamma\left(\frac{\nu}{\alpha}\right) }{\Gamma\left(\frac{1}{2}-i\frac{\omega}{2\alpha}+\frac{1}{4}\left(\nu+\zeta~\right)
 \right) \Gamma\left(\frac{1}{2}-i\frac{\omega}{2\alpha}+\frac{1}{4}\left(\nu-\zeta~\right)
 \right)}.
\end{equation}
For $\nu > 1$, we notice that the power-law term $(\alpha x)^{\frac{1}{4}(1 - \nu)}$ in (\ref{lim}) diverges as one approaches infinity (which corresponds to $\alpha x\rightarrow 0^{-}$), while the other term decays towards zero. However, the presence of poles in the Gamma function at negative integers may render this problematic term effectively vanish. As a result, we derive a discrete set of countable frequencies that characterize the QNM solutions, 
\begin{equation}
\label{f3}
\omega_\pm = -i\alpha \left( 2n + 1 + \frac{1}{2} (\nu\pm \zeta)\right),
\end{equation}
with $n\in \mathbb{Z}_{\ge 0}$. 
These results, as shown in (\ref{f3}), are consistent with those presented in \cite{Bin}, \cite{molina}, and \cite{nos}. Besides,  Eq. (\ref{f3}) can be derived by analyzing the singular points in the transfer matrix—or transmission coefficient—where $\mathbb{T}(\omega_{\pm}) = \infty$. This approach was previously demonstrated in the context of the P\"oschl-Teller potential \cite{cevik} but also in the case of a generalized P\"oschl-Teller potential \cite{silva}. It should be mentioned that there are other kinds of interesting situations that were analyzed in \cite{nos}, such as: 
\begin{enumerate}[i.]
    \item QNMs with the usual exponentially suppressed oscillatory
behavior for $V_{0}>0$ and $V_{1}>\alpha^{2}/4$,
\item the so-called algebraically special QNMs  for  $V_{1}\leq \alpha^{2}/4$, and
\item unstable modes for small $V_{1}/\alpha^{2}$.
\end{enumerate}
 For more information on these possibilities, the reader may  consult Ref. \cite{nos}.

The QNM solutions has the following effective boundary condition at $x=0$,
\begin{equation}
\label{bc}
\lim_{x\to 0^-} (\alpha x)^{-\kappa} \left[(\alpha x)^{3\over4} u'_{I}(x) -
\frac{1}{(\alpha x)^{1\over4}} \alpha\left(\frac{1}{4} +\kappa \right)   u_{I}(x)\right] = 0,
\end{equation}
where $\kappa=\sqrt{\frac{1}{16}+\frac{V_0}{\alpha^2}~}>0$. Eq. (\ref{bc}) resembles the condition reported in \cite{nos}. Another interesting point is to examine whether  or not the functional energy remains bounded spatially  for the QNMs solution at infinity \cite{nos}. 
As long as $\kappa>7/4$, the functional energy converges to zero as $\alpha x \rightarrow 0^{-}$. 

Now we  are in position to discuss  the role plays by the analytical continuation of the QNM problem in the case of the BTZ black hole. We will give a proof of concept by analyzing one case based on the ideas presented in Sec. II.  The outcome of applying the analytical continuation, defined as $x=iy$, to the QNMs of the BTZ black hole \cite{borel} is as follows.  The solution   $u_{I}(x)$ associated with the potential $V(x)$ will transform into quantum eigenstates $\psi=u_{I}(V\rightarrow -V(iy, \alpha'), \omega\rightarrow -i\omega)$  of the inverted potential barrier, $\tilde{V}=-V$.  Thus the Schr\"odinger equation becomes, 
\begin{equation}\label{dual}
\Big(-\frac{d^2}{dy^2} - \frac{V_{0}}{\sinh^{2}(\alpha' y)}- \frac{V_{1}}{\cosh^{2}(\alpha' y)}\Big)\psi=E \psi.
\end{equation}
It is important to stress that $\alpha$ parameter  must accommodate the modification introduced by the analytic continuation in order to keep  the shape of potential unspoiled \cite{ba3}. As result of that procedure, the energy eigenvalue ($E=-\omega^{2}$) now  reads,
\begin{equation}
\label{E2}
E = -\alpha'^{2} \left( 2n + 1 + \frac{1}{2} (\nu\pm \zeta)\right)^{2}.
\end{equation}
Including  these transformations in the definitions of $\nu$ and $\zeta$, the combination appearing in (\ref{E2}) becomes, $\nu\pm \zeta=\sqrt{1-4\frac{V_0}{\alpha'^2}}\pm \sqrt{1+4\frac{V_1}{\alpha'^2}}$. The latter fact pinpoints a potential issues regarding the self-adjoint property of the  Schr\"odinger operator presented in (\ref{dual}) provided the energy can take complex value. The reason for suspecting that something might have gone wrong around $y=0$ can be easily  confirmed by  expanding the inverted potential around that point. The leading term is  $\tilde{V}=-V_{0}/(\alpha' y)^{2}<0$. This kind of potential yields non-self-adjoint operator on a Hilbert space $\mathbb{L}^2[(-\infty, 0), dy]$ \cite{grif1}, \cite{fulop}.  

From now we will focus on the properties of  the Schr\"odinger operator (\ref{dual}) and the effective boundary condition around $y=0$. To do so, we follow a well-established protocol based on the Von Neumann's theorem  \cite{GTV}, \cite{AJP}. We begin by computing the subspace of solutions  with purely imaginary eigenvalues denoted as $N_\pm = \left\{ \phi\in  D({\mathcal{H}^\dagger}), \quad {\mathcal{H}}\phi = \pm i \phi\right\}$ \cite{GTV}, where $\mathcal{H}$  stands for  the Schr\"odinger operator presented in  (\ref{dual}) . In our case, near $y=0$ these solutions are given by  
\begin{equation}
\label{phis}
\phi_{\pm} =  (\alpha' y)^{1/4}\left(  A_{\pm} (\alpha' y)^{ \bar{\kappa}} + B_{\pm} (\alpha' y)^{- \bar{\kappa}}  \right). 
\end{equation}
Here $\bar{\kappa}=\kappa(V_{0}\rightarrow -V_{0}, \alpha \rightarrow \alpha')$. Eq.  (\ref{phis}) indicates that, locally, in each case $\pm$, only one of the solutions is square-integrable with respect to the measure $dy$. This fact shows that the dimension of the subspaces $N_\pm$ is at least one in both cases, and consequently, the operator admits a self-adjoint extension parametrized  by the $U(1)$ group. In other words, there are an infinite number of self-adjoint extensions which can be written as $\phi=\phi_{+} + s \phi_{-}$ with $s \in \mathbb{C}$ such that $|s|=1$. For any element $\psi \in D(\mathcal{H}^\dagger)$, in order to ensure that the  self-adjoint extensions are  well-defined, they must fulfill the following boundary condition,
\begin{equation} \label{ext} \left\langle \phi, \mathcal{H} \psi \right\rangle - \left\langle \mathcal{H} \phi, \psi \right\rangle = \lim_{y \to 0^-} \left[ \bar{\phi}(y) \psi'(y) - \bar{\phi}'(y) \psi(y) \right] = 0, 
\end{equation}
where the bracket $\left\langle, \right\rangle $  refers to the usual inner product in  $\mathbb{L}{}^2\big([-\infty,0), dy\big)$. For the sake of simplicity, let us corroborate whether the  analytically continued eigenstates satisfy the same effective boundary condition of the QNMs (\ref{bc}) or not. We only consider the situation associated with the QNMs, so from the general combination the $A_{\pm}$ terms must be omitted, while the identification $u=\psi$ is made explicit. To keep things simple, we consider the case in which $\bar{\kappa} \in \mathbb{R}$, thus $0<V_{0}/\alpha'^{2}<1/4$ \cite{nos}.  The boundary condition
(\ref{ext}) can be recast as
\begin{equation}
\label{bcmof}
\lim_{y\to 0^-} (\alpha' y)^{-\bar{\kappa}} \left[ (\alpha' y)^{3\over4} u'(y) -
\frac{1}{(\alpha y)^{1\over4}} \alpha'\left(\frac{1}{4} -\bar{\kappa} \right)   u(y)\right] = 0. 
\end{equation}
The physical implications derived from equation (\ref{bcmof}) can be summarized as follows. Upon determining the self-adjointness of the generalized (inverted) P\"oschl-Teller operator as described in (\ref{dual}), and imposing the necessary conditions for self-adjointness at the boundary $y=0$, we find that the effective boundary conditions associated with the quasinormal modes  differ from the original conditions presented in (\ref{bc}). Specifically, for the range $0 < \frac{V_{0}}{\alpha'^{2}} < \frac{1}{4}$, the self-adjoint extensions do not fulfill to the same boundary condition specified in (\ref{bc}). This indicates that the analytically continued QNMs do not belong within the domain of any self-adjoint extension \cite{nos}. This observation further supports our conclusions regarding the analytical continuation method and the (inverted) P\"oschl-Teller potential, as presented in Sec. II.

\section{Summary} 
We discussed the issues that emerge when employing the analytical continuation method to obtain the complete set of quasinormal modes in solvable scenarios, including the P\"oschl-Teller potential and the BTZ black hole case. The absence of (essentially) self-adjointness in the Schr\"odinger operator with the inverted potential significantly restricts the viability of this approach \cite{nos}.  Nevertheless, it would be interesting to revisit this BTZ case in the light of the recent developments for the pseudospectrum of the P\"oschl-Teller operator \cite{pseudo0}, \cite{pseudo1} and in the case where the black hole is asymptotically AdS \cite{pseudo2}, \cite{pseudo3}, \cite{pseudo4}, \cite{pseudo5}. The latter point will be addressed elsewhere.


\acknowledgments

J.C.F. is supported by Conselho Nacional de Desenvolvimento Cient\'ifico e Tecnol\'ogico (CNPq, Brazil) and  Funda\c{c}\~ao de Amparo \'a Pesquisa e Inova\c{c}\~ao Esp\'irito Santo (FAPES, Brazil). A.S. is partially supported by Conselho Nacional de Desenvolvimento Cient\'ifico e Tecnol\'ogico (CNPq, Brazil).

\end{document}